\documentclass[pdflatex,sn-mathphys-num]{sn-jnl}

\usepackage{array}
\usepackage{longtable}
\usepackage{multirow}
\usepackage{float}
\usepackage[table,xcdraw]{xcolor}
\usepackage{graphicx}%
\usepackage{amsmath,amssymb,amsfonts}%
\usepackage{amsthm}%
\usepackage{mathrsfs}%
\usepackage{accessibility}
\usepackage[title]{appendix}%
\usepackage{xcolor}%
\usepackage{textcomp}%
\usepackage{manyfoot}%
\usepackage{booktabs}%
\usepackage{algorithm}%
\usepackage{algorithmicx}%
\usepackage{algpseudocode}%
\usepackage{listings}%
\usepackage{fix-cm}

\theoremstyle{thmstyleone}%
\theoremstyle{thmstyletwo}%

\theoremstyle{thmstylethree}%

\begin{document}

\title{Variable Selection in the Context of AI Fairness}

\author*[1]{\fnm{Ivan Luciano} \sur{Danesi}}\email{ivanluciano.danesi@unicredit.eu}
\equalcont{These authors contributed equally to this work.}
\author[2]{\fnm{Chiara} \sur{Frigerio}}\email{chiara.frigerio@unicatt.it}
\equalcont{These authors contributed equally to this work.}
\author[2,3]{\fnm{Fabio} \sur{Maccaferri}}\email{fabio.maccaferri@unicatt.it}
\equalcont{These authors contributed equally to this work.}
\author[2,3]{\fnm{Giorgio Alessandro} \sur{Motta}}\email{giorgioalessandro.motta@unicatt.it}
\equalcont{These authors contributed equally to this work.}
\author[3]{\fnm{Pietro} \sur{Zecca}}\email{pietro.zecca@cetif.eu}
\equalcont{These authors contributed equally to this work.}

\affil[1]{ \orgname{UniCredit S.p.A.}, \orgaddress{\city{Milan}, \country{Italy}}}

\affil[2]{\orgname{Universit\`{a} Cattolica del Sacro Cuore}, \orgaddress{\city{Milan}, \country{Italy}}}

\affil[3]{\orgname{Cetif}, \orgaddress{\city{Milan}, \country{Italy}}}

\abstract{
{\bf Background:} 
    Fairness in AI systems has become more important with recent regulatory demands, such as the EU AI Act. Traditional approaches often do not take into account philosophical ethics and social awareness. Variable selection processes, in particular, can introduce implicit bias, affecting equity across different subgroups.
    
    {\bf Objectives:}
    We discuss a mathematical approach that evaluates fairness in AI, aligning mathematical methodologies with ethical considerations and regulatory requirements. Our aim is to advocate for interdisciplinary collaboration to address fairness, emphasizing the importance of understanding broader ethical and societal contexts.
    
    {\bf Methods:}
     Our approach emphasizes maintaining all potentially relevant variables to allow for more granular fairness assessments and to reduce implicit bias.

    {\bf Results:}
    The findings suggest that the exclusion of sensitive or critical variables may compromise equity between subgroups. In contrast, retaining all relevant variables could reduce implicit bias. Thus, the interdisciplinary approach could provide deeper insight into the ethical implications and compliance with regulatory standards.
    
    {\bf Conclusions:}
     By integrating a mathematical approach with ethical and social awareness, we suggest more equitable outcomes and responsible AI deployment. This work underscores the necessity of interdisciplinary collaboration in effectively addressing fairness in AI systems aligned with the objectives of the European Union’s AI Act, which seeks to promote trustworthy and fair AI systems.
}

\keywords{Bias, European AI Act, Fairness in Machine Learning, Variables Selection}

\maketitle

\section{Introduction} \label{sec:Introduction}

The concept of fairness in machine learning and artificial intelligence has become more important as these technologies permeate various aspects of society. However, the notion of fairness is multifaceted and often context-dependent, making it challenging to define and measure universally \cite{verma2018, cheng2020}. This complexity is further compounded when we consider that our models, built on simplified representations of reality through data, may not fully capture the intricacies of the real world \cite{benthall2019}.  Our contributions are as follows:

\begin{enumerate}
    \item We discuss current efforts to understand the broader implications of fairness beyond mathematical and statistical analysis.
    \item We bridge theory and practice by discussing fairness auditing and compliance with international standards (e.g., the EU AI Act).
    \item We discuss a mathematical approach illustrating the conditions under which retaining all variables, including sensitive ones, may lead to fairer classification outcomes.
\end{enumerate}

We do not attempt to prescribe which definitions of fairness are appropriate for each particular situation. Instead, we explore the implications of including all variables in our models, rather than excluding potentially sensitive attributes. This approach aligns with recent work by Scutari et al. \cite{scutari2022}, who propose achieving fairness through simple regularization techniques rather than variable exclusion.
We argue that by retaining all variables, we may be able to capture more nuanced relationships in the data, potentially leading to models that are not only more accurate, but also more equitable when viewed through multiple fairness lenses \cite{foulds2020}. However, this approach is not without challenges. As Li et al. \cite{li2022} discuss in their work on trustworthy AI, we must carefully consider the ethical implications of using sensitive attributes, even when done with the intention of improving fairness.
By examining these issues, we hope to contribute to the ongoing dialogue on fairness in AI, as outlined in comprehensive reviews by González-Sendino et al. \cite{gonzalez2023} and Chen et al. \cite{chen2023}. Our goal is to provide insights that can help researchers and practitioners navigate the complex landscape of AI fairness, moving toward models that are not only statistically sound but also aligned with broader societal values of equity and justice.

Firstly, we discuss the challenges of defining fairness in AI, highlighting recent regulatory developments, especially the GDPR and the AI Act, and highlight the importance of aligning technical methods with broader ethical and social values, and give an overview of the methodology used in the paper (Section  \ref{sec:Introduction}). Secondly, we further discuss that bias and fairness are complex and context-dependent, and fair AI requires multidisciplinary approach throughout integrating ethical reflection alongside technical solutions to promote equitable outcomes (Section  \ref{sec:Unawareness}). Finally, we present a mathematical framework to discuss that excluding variables in AI models can lead to a significant loss of fairness and explainability (Section \ref{sec:BalancingAct}).

\subsection{Fairness: The Bias behind the Variable Choosing} \label{sec:The Bias behind the Variable Choosing}

In this subsection, we illustrate how previous works already discuss that variable exclusion does not necessarily guarantee fairness and may, in practice, propagate subtle forms of discrimination. We illustrate, through case studies, that there is an ongoing sentiment that there is a need to ensure that fairness within model building.

The concept of fairness extends beyond the one we can explain with statical or mathematical formula, As Verma and Rubin \cite{verma2018} point out, the question ``Is the classifier fair?'' is deceptively simple. The answer depends critically on which definition of fairness one adopts. There are numerous statistical measures of fairness, each capturing different aspects of equity and nondiscrimination. These include demographic parity, equal opportunity, and individual fairness, among others \cite{xivuri2021, lohia2018}. However, it is essential to recognize that these measures are ultimately calculations of error distributions and do not necessarily reflect the full complexity of societal fairness \cite{zliobaite2017}.

Moreover, the simplification of reality through data collection and model building processes can lead to inadvertent biases. Benthall and Haynes \cite{benthall2019} argue that even the categories we use in machine learning, particularly racial categories, can embed historical biases and oversimplify complex social constructs. This simplification can result in models that, while mathematically "fair" or correct according to certain metrics, may still perpetuate or exacerbate existing societal inequities \cite{weinberg2022}.

It is well-documented that eliminating sensitive variables such as race, gender, or socioeconomic status from AI models does not necessarily prevent discriminatory outcomes \cite{zliobaite2017}. Recent studies have confirmed that seemingly neutral features—so-called \emph{proxy variables}—can encode and perpetuate social biases due to high correlation with protected attributes~\cite{ferrara2023, cornacchia2023, cerqueira2022}. 

For example, in credit scoring, the omission of explicit race information can appear to satisfy fairness mandates; however, other inputs such as postal code, employment history, or consumer spending patterns frequently act as effective proxies for race and socioeconomic background~\cite{cerqueira2022}. This practice can result in disguised bias, in which models indirectly disadvantage certain groups despite the removal of sensitive variables.

In healthcare, Cerqueira et al.~\cite{cerqueira2022} demonstrated that risk prediction algorithms using healthcare spending as a feature undermined fairness, since spending often reflected underlying socioeconomic disparities even when race was omitted from the model. Similarly, Cornacchia~\cite{cornacchia2023} used counterfactual reasoning to show that discrimination can persist via non-linear interactions among non-sensitive variables. In the employment domain, audits reveal that gender is indirectly reintroduced in hiring models when educational institutions or extracurricular activities, which are correlated with gender, remain among the predictors.

These cases highlight the insufficiency of ``fairness through unawareness’’ and underscore the necessity for vigilant auditing of both direct and indirect sources of bias in contemporary AI systems~\cite{ferrara2023, gichoya2023}.

\subsection{AI Fairness, European Regulatory Landscape} \label{sec:AI Fairness, European Regulatory Landscape}

As mathematical models do not provide sufficient element to protect and ensure sufficient fairness through ignorance all over the world, the regulatory body drafts a regulation to protect the consumer from the risk of AI and inappropriate data use.

In Europe, the first step was with the General Data Protection Regulation (GDPR), which is a comprehensive data protection law implemented by the European Union that establishes strict guidelines for the collection and processing of personal information. In force since May 2018, GDPR aims to enhance individuals' control over their personal data, ensuring privacy and transparency in data handling \cite{EUGDPR}. The GDPR also imposes several obligations on AI systems that process personal data. Specifically, it requires AI developers to ensure lawful processing, transparency, and accountability in data usage. This means that any AI system that uses personal data must have a clear legal basis, often requiring the explicit consent of the individuals whose data are being processed \cite{WrightCrootof}. The regulation emphasizes the need for data minimization, requiring organizations to collect only the data necessary for their stated purposes, directly affecting AI training and functionality. 

A more recent initiative is the European AI Act, which creates a regulatory framework for Artificial Intelligence (AI) systems deployed within its jurisdiction. The Act aims to safeguard fundamental rights and ensure safety and transparency, respecting privacy and data protection rules. It underscores the importance of high standards in the quality and integrity of processed data and mandates transparency in AI operations, facilitating traceability and explainability \cite{EUAIAct2024}. By emphasizing diversity, nondiscrimination, and fairness, the Act requires AI systems to avoid discriminatory impacts and unfair biases.
As The AI Act underscores a firm commitment to safeguarding human autonomy and decision making by prohibiting AI practices that manipulate behavior or exploit vulnerabilities \cite{EUAIAct2024}. It strictly bans AI systems designed for social scoring and those that classify individuals in potentially discriminatory or harmful ways, thus upholding human dignity and fundamental rights \cite{EUAIAct2024}. The Act emphasizes the importance of accuracy, robustness, and cybersecurity in high-risk AI systems \cite{EUAIAct2024}, fostering a convergence of correctness and fairness. 
Europe is not the only one moving on with AI regulation; the Americas and Asia are also progressing with their own regulatory frameworks, as shown in Table 1 \ref{tab:comparison_matrix}

\begin{longtable}{|>{\raggedright}p{2,5cm}|>{\raggedright}p{2,5cm}|>{\raggedright}p{2,5cm}|>{\raggedright}p{2,5cm}|>{\raggedright\arraybackslash}p{2,5cm}|}
\hline
\textbf{Criteria} & \textbf{European AI Act} & \textbf{GDPR (EU)} & \textbf{USA Regulation} & \textbf{Asian Regulation} \\ \hline
\endfirsthead
\hline
\textbf{Criteria} & \textbf{European AI Act} & \textbf{GDPR (EU)} & \textbf{USA Regulation} & \textbf{Asian Regulation} \\ \hline
\endhead
\multicolumn{5}{|r|}{{Continued on next page}} \\ \hline
\endfoot
\endlastfoot

\textbf{Scope} & AI systems deployed in the EU, with a risk-based approach. & Processing of personal data within the EU. & Varies significantly; sector-specific like HIPAA for health, FCRA for consumer credit. & Range from broad data protection (e.g., Japan's APPI, Singapore's PDPA) to more fragmented frameworks. \\ \hline
\textbf{Primary Focus} & Safety, transparency, and fundamental rights in AI use. & Protection of personal data and privacy rights. & Sectoral, focusing on consumer protection, privacy, and data breach notifications. & Varies: from data protection to specific AI guidelines (e.g., China's AI development plans, Japan's AI strategy). \\ \hline
\textbf{Fairness and Bias} & Requires measures against unfair bias in high-risk AI systems. & Implicit in the principles of lawful, fair, and transparent processing. & Fairness considered under certain laws (e.g., ECOA in lending). & Emerging focus on AI ethics and fairness, but less codified in law compared to EU/USA. \\ \hline
\textbf{Enforcement} & National supervisory authorities; significant fines for non-compliance. & National Data Protection Authorities; fines up to 4\% of annual global turnover or €20M. & Varies by law and enforcing body (FTC, state attorneys general). & Varies widely; Japan and Singapore show strong enforcement, while others are evolving. \\ \hline
\textbf{Global Impact and Influence} & Likely to set a global standard for AI regulation. & Has set a global standard for data protection laws. & Influential, particularly in areas without strong local regulations; California Consumer Privacy Act (CCPA) seen as benchmark. & Increasing influence, especially in regions developing their own AI and data protection standards. \\ 
\hline

\caption{Comparison Matrix of AI and Privacy Regulations}
\label{tab:comparison_matrix}
\cite{EUAIAct2024,EUGDPR,AILawChinaEUUS,GDPRimpactonAI,GlobalAIRegulations,EUUSAIRegulationComparison, WrightCrootof}
\end{longtable}

According to our paper, the Act points out that AI systems must be fair to be correct, recognizing inherent biases in AI models and mandating measures for their identification and mitigation to ensure that fairness is not sacrificed for correctness\cite{bellamy2018}. We can find a link between GDPR and the AI Act, both frameworks aim to protect individuals' rights while fostering responsible technological advancement. The AI Act builds on the principles laid out in GDPR, particularly in terms of transparency and accountability, establishing a regulatory paradigm that seeks to balance innovation with ethical considerations in AI deployment \cite{GDPRimpactonAI}.

\subsection{Methodology} \label{sec:Methodology}
\begin{figure}[ht]
    \centering
    \includegraphics[width=1\textwidth]{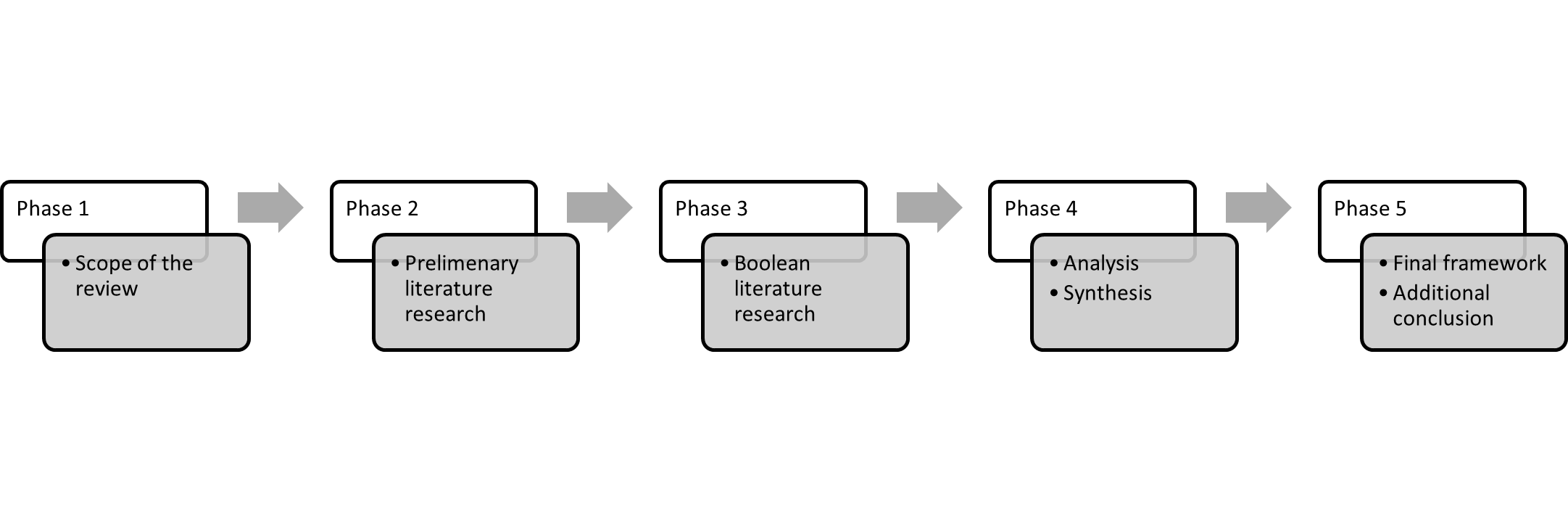}
    \caption{The five sequential phases in the research design of this project}
    \label{fig:sequentialphases}
\end{figure}

As this paper presents a way to assess fairness in AI, emphasizing the importance of comprehensive variable selection, we investigate the application of static fairness for AI modeling by constructing a general mathematical framework.
Therefore, a pre-literature review was conducted to identify the meaning of fairness in AI by looking at the main digital libraries such as Elsevier, Google Scholar and Scopus, as shown in Figure \ref{fig:sequentialphases}.

\begin{figure}[H]
    \centering
    \includegraphics[width=0.6\textwidth,height=0.6\textheight,keepaspectratio]{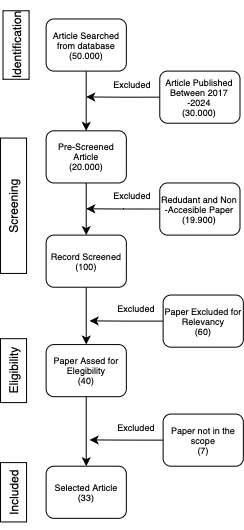}
    \caption{Preliminary literature review workflow leading to the identification of the papers used to define the main AI fairness dimensions discussed in this study.}
    \label{fig:classification}
\end{figure}

This initial research, conducted using the four strategies of Rolwey and Slack \cite{rowley2004} led to more than 50.000 results using keywords such as \textit{AI ethics, AI fairness metrics, Explainable AI or XAI, AI transparency, Robust AI, Fairness in AI-driven decision making, Sustainable AI in business ecosystem, AI Act EU, Federated learning for privacy, Differential privacy in AI, AI auditing frameworks, Discrimination, Trade-offs in fair AI, social impact of AI,ethics and AI.} A further selection of articles was performed using the recommendations of \cite{rowley2004} and \cite{kitchenham2007}, namely that the articles should: 1) be relevant to the research subject, 2) be up to date, 3) have extensive sources referencing and 4) be written by an authoritative author. From the pre-literature review papers, we extracted the elements of AI Fairness for a total of 33 papers, as shown in Figure \ref{fig:classification} 

The final set of 33 selected papers constitutes the literature foundation of this work and is referenced throughout the remainder of the paper to discuss the main dimensions of AI fairness, including bias, explainability, transparency, robustness, and regulatory compliance. Afterwards, the hypothesis and the research question were formulated.\\
\textbf{Hypothesis}: Let D be the training data set of a classifier C, consistent, representing the available operational reality as completely as possible.\\
\textbf{Research Question}: Given a data set $D$, the fairness assessment of the classifier $c$ can be addressed with a higher level of reliability awareness as the variables of the learning data set $D'$ $(D' \subseteq D)$ potentially pertinent to the purpose of the application of the classifier increase without \textit{a priori} screening and elimination actions. To maximize fairness, we recommend an initial training phase with the full dataset 
\textit{D}, ensuring the model incorporates all observed data distributions. This practice helps mitigate the risk of overlooking minority group patterns.

\section{Unawareness as an approach to Fairness: A Blessing in Disguise} \label{sec:Unawareness}

The current development and deployment of AI technologies have raised ethical, fairness, and privacy concerns. Among the efforts to address these concerns, one prominent approach has been the establishment of guidelines and frameworks for trustworthy AI \cite{li2022}.Trustworthy AI encompasses principles such as robustness, explainability, and fairness, in order to ensure that AI systems are developed and operated ethically justifiable and respect user privacy \cite{EUAIAct2024,EUGDPR}.
These principles are not isolated; instead, they are part of a network of technical, ethical, and other requirements. Therefore, asking if our model can be fair or unfair is difficult, as fairness is an ethical construct that integrates with multiple aspects of AI development and operation. This interconnectedness emphasizes the need for a holistic evaluation of all aspects to ensure that human values are upheld in AI systems. Here, we focus on a broader sense of ethical considerations, acknowledging that fairness cannot be assessed in isolation but must be understood within the complex ecosystem of AI requirements and principles.
Bias is typically understood as a systematic error that leads to skewed results, often manifesting as prejudice against certain groups. González-Sendino et al. (2023) emphasize that bias can originate from a variety of sources, including biased training data and model architectures \cite{gonzalez2023}.
Fairness, even if it is related to bias, involves normative principles related to justice and equity in outcomes. Verma and Rubin (2018) provide insight into various fairness definitions, showing that fairness encompasses a spectrum of approaches and philosophical viewpoints that extend beyond simple bias correction \cite{verma2018}.
Addressing bias often means correcting data imbalances or adjusting algorithms to minimize prejudicial impacts. Techniques such as those discussed by Bechavod and Ligett (2018) penalize unfair classification, aiming to identify and rectify specific biases \cite{bechavod2018}.
Fairness requires a broader ethical consideration, as explored by Li et al. (2022), who argue that trustworthy AI must embody principles such as transparency and accountability \cite{li2022}. This involves implementing fair practices throughout the AI lifecycle, not just correcting bias post hoc.
Focusing solely on bias can lead to reactive measures that might fix immediate issues but overlook the broader context of equity. For instance, Žliobaitė (2017) discusses the importance of measuring discrimination to understand bias but underlines that these measurements alone do not guarantee fair outcomes \cite{zliobaite2017}.
Fairness, as explained by Richardson and Gilbert (2021), involves constructing systems that proactively ensure equitable treatment and outcomes, which can span across various models, use cases, and societal impacts \cite{richardson2021}.
Krchova et al. (2023) highlight that while bias may focus on specific attributes, fairness often necessitates an intersectional approach—considering the nuanced interplay between multiple attributes and identities within algorithmic contexts \cite{krchova2023}.
The quest for fairness in artificial intelligence (AI) systems has become important as these technologies influence crucial aspects of societal functions, including hiring, financial risk assessment, and personal identification. Moreover, definitions of fairness vary, from ensuring that AI systems do not perpetuate bias or discrimination \cite{gonzalez2023} to more formal mathematical formulations aimed at quantifying and mitigating unfairness in algorithmic decision-making \cite{verma2018}. The challenge lies not only in defining fairness but also in operationalizing these definitions in diverse contexts \cite{bechavod2018,krchova2023}.
Defining fairness within AI frameworks embodies a complex endeavor, presenting multiple definitions that attempt to capture equity in algorithmic decision-making. These encompass more ways to approach fairness in terms what should be computed \cite{benthall2019}:
\begin{itemize}
\item \textbf {Fairness through Unawareness}: An algorithm is fair as long as protected attributes $A$ are not explicitly used in decision making. \cite{dwork2012fairness}

\medskip

\item \textbf{Demographic Parity} A predictor $\hat{Y}$ satisfies demographic parity if 
$$
P(\hat{Y} \mid A = 0) = P(\hat{Y} \mid A = 1)
$$

\medskip

\item \textbf{Equality of Opportunity} 
A predictor $\hat{Y}$ satisfies equality of opportunity if 
$$
P(\hat{Y} = 1 \mid A = 0, Y = 1) = P(\hat{Y} = 1 \mid A = 1, Y = 1)
$$
\end{itemize}
Instead, if we look for fairness based on group or individual, we have:
\begin{itemize}
    \item individual fairness, advocates for consistent treatment of similar individuals \cite{lohia2018,krasanakis2024};
    \item group fairness, seeks to eliminate discrimination based on demographic factors \cite{lohia2018,krasanakis2024};
    \item subgroup fairness,scrutinizes fairness at the intersectionality of multiple protected attributes, ensuring fairness across a broader spectrum of identity group combinations.
\end{itemize}
Each offering a different lens through which fairness can be evaluated and achieved \cite{varona2022,richardson2021,chen2023,gonzalez2023,li2022}.

The significance of these fairness perspectives is amplified by the potential social ramifications of systematic unfairness perpetuated by AI systems. Therefore, to bring fairness in AI is a difficult journey, connected with the complexities of real-world inequities. AI models serve as artificial constructs through which we attempt to impose equity and justice. In some cases, this requires the removal of certain variables to prevent discrimination. However, this approach can lead to a suboptimal system, where the exclusion of key variables detracts from the model's accuracy and effectiveness. It is crucial to recognize that achieving fairness is not merely a technical challenge, but also an ethical obligation. This requires a holistic approach, integrating fairness throughout the AI development lifecycle to build systems that not only mirror societal values but actively enhance them, promoting equitable outcomes without compromising performance. Unfair biases, whether originating from data, models, or procedural practices, can result in systemic disadvantages for underprivileged groups, eroding trust in AI technologies and hindering their positive impact on society. 
Therefore, by intentionally excluding sensitive attributes from the decision-making process, we limit the interpretability of the Fairness. In the following chapter, we will dive into the formal definition of this concept and explore how it can be applied in practice to develop fairer AI models.

\section{Balancing Act: Optimizing Fairness and Error in AI Classification Models Through Comprehensive Variable Inclusion}\label{sec:BalancingAct}

We explore the intricate trade-offs between fairness and predictive accuracy in AI classification models, emphasizing the importance of comprehensive inclusion of variables. We analyze how excluding sensitive attributes, often an intuitive approach to mitigate bias, can paradoxically diminish both the fairness and the explanatory power of machine learning systems. By systematically evaluating model performance and fairness through varying feature subsets, we illustrate the theoretical implications of different variable selection strategies. Ultimately, we advocate for a nuanced methodology that first optimizes classification error across all available predictors and subsequently rigorously assesses and addresses fairness. This integrated approach seeks to achieve models that are not only statistically fair but also aligned with ethical and societal expectations in real-world applications.

As discussed in Section 2, a classifier \textit{C} is said to satisfy fairness through unawareness if the prediction function excludes sensitive attributes from its input. Let \( X \) denote the full set of input variables and \( X' \subseteq X \) the subset of sensitive (or protected) attributes, such as gender, race, or age. Under fairness through unawareness, the classifier operates only on the non-sensitive features, \textit{i.e.},
\[
\hat{Y} = f(X \setminus X'),
\]
where the sensitive attributes in \( X' \) are explicitly excluded from the prediction process.

It is not necessarily the case that reducing the number of input variables has a negligible impact on the resulting model. For instance, consider the following scenario: a subject is randomly selected for analysis, and a synthetic dataset of 100,000 entries is generated to ensure statistical robustness. All characteristics remain constant except for the sensitive attribute under examination. The classification algorithm is then applied to this dataset to generate predictions. If the predictions remain identical (or nearly identical) in all records, regardless of variations in the sensitive attribute, it can reasonably be inferred that this variable does not significantly influence the model output. As a result, the variable could be excluded from the training process without compromising predictive performance.

However, there exist numerous situations where the removal of certain variables does, in fact, affect the model’s behavior, and such exclusions should be carefully evaluated in context. This consideration provides the foundation for the discussions that follow. In what follows, we assume that unfairness arises as a consequence of model bias.\\

Let $D$ be an optimized training sample for a generic AI model $M$. 
Let $m,n \in \mathbb{N}, n\geq m$ denote respectively the number of predictors 
in $D$ and in a subset $D'\subset D$ obtained by the removal of exactly 
$n-m$ variables.\\
To train model $M$ on $D$, one can easily adopt a step forward feature selection:
$\forall k \in \mathbb{N}$, so that $0\leq k \leq m$, however $k$ predictors 
are chosen we can use them to estimate model parameters generating a finite sequence 
of models $M_k$.\\
Error optimization (\textit{e.g.} MSE) allows to choose a subset of cardinality $\bar k$ so that 
$M_{\bar k, \bar X_{\bar k}}(X_1,...,X_m)$ makes the smallest mistake $\varepsilon_{\bar k}$.\\
Since $D'$, is arbitrarily chosen among the subsets of assigned cardinality $n\leq m$ in $D$ , 
we can assume wlog that $\varepsilon_{n}\geq\varepsilon_{\bar k}$.\\ \\

Let $X$ be the set $\{x_1, \ldots, x_n\}$ of variables vectors $V = \{v_1, \ldots, v_p\}$ representing the population of $p$ subjects on which the classifier $C$ has been trained.  
Thus each vector $x_i$ is defined as $x_i = \{v_{i,1}, \ldots, v_{i,p}\}$.

The classifier $C$ has determined the best error on the matrix:

\[
\left(
\begin{array}{ccc}
v_{1,1} & \cdots & v_{1,p} \\
v_{2,1} & \cdots & v_{2,p} \\
\vdots & \ddots & \vdots \\
v_{n,1} & \cdots & v_{n,p}
\end{array}
\right)
\]

where rows represent each subject in the given population.

Let $S$ be the set of variables identified as "sensitive", such that $S \subset V$, then $ V \setminus S = V' = \{v_1, \ldots, v_k\} \setminus \{s_1, \ldots, s_t\} = \{v_1, \ldots, v_{p-t}\} $.

Training the classifier on $ V' $ is not optimal, since we have higher values in  the error matrix.  
We claim that, in addition to sub-optimality in term of performance, excluding sensitive variables during training is counterproductive also because the information loss may affect fairness as well.

Let $M, M'$ be the outcome of the classifier $C$ trained on $V, V'$  respectively.
Let us say that $\Theta$ is the family of all possible fairness definitions, and assume that $\Delta$ identifies a specific class of subjects potentially discriminated by classification choices. Assessing fairness of a classifier $C$ is equivalent to defining a real valued function $f_{\theta,\Delta}$ for any $\theta\in\Theta$ and $\Delta$, depending on the sample and on the training of the classifier $C$.\\ 
Since training $M$ is optimal, any other choice including $M'$ is suboptimal in terms of error. In other words, reducing the amount of significant information available results in model training that can introduce an higher level of bias.\\
Now, this bias affects classification in three possible way:
\begin{enumerate}
\item The trained model M' is less fair than model M. $$f_{\theta,\Delta}(V,M) \geq f_{\theta,\Delta}(V',M')$$
\item The trained model M' is as fair as model M. $$f_{\theta,\Delta}(V,M) = f_{\theta,\Delta}(V',M')$$
\item The trained model M' is more fair than model M. $$f_{\theta,\Delta}(V,M) \leq f_{\theta,\Delta}(V',M')$$
\end{enumerate}
as shown in the table below, we may conclude that the model $M$ trained on $V$ (all variables) is usually preferable or, at worst, equivalent (2 of 3 conditions). This justifies the recommendation to train the model on all variables in order to optimize predictive performance before evaluating fairness by identifying the best trade-off between fairness and prediction error (Table \ref{fairness-table}), since excluding variables may negatively affect classification performance.
\renewcommand{\arraystretch}{1.2} 

\renewcommand{\arraystretch}{1.5} 
\setlength{\tabcolsep}{5pt}       
\begin{table}[h!]
\caption{Trained Model Choice based on fairness level}
\label{fairness-table}
\centering
\begin{tabular}{|c|c|}
\hline
\rowcolor[gray]{0.9} \textbf{Cases} & \textbf{Trained Model Choice based on fairness level} \\
\hline
$f_{\theta,\Delta}(V,M) \geq f_{\theta,\Delta}(V',M')$ & $M$ \\
\hline
$f_{\theta,\Delta}(V,M) = f_{\theta,\Delta}(V',M')$ & $M$ \\
\hline
$f_{\theta,\Delta}(V,M) \leq f_{\theta,\Delta}(V',M')$ & $M'$ \\
\hline
\end{tabular}

\end{table}

Moreover, if we assume that we are in the third scenario, knowledge of the sensitive features allows us to identify $i$ so that $v_i=\{v_{1,i},\cdots\,v_{n,i}\}\in \Delta$ is fairly classified by $M'$ and not by $M$.
Lets define a new outcome for the classifier $C$ as follows:
\[
\tilde{M}_i(v_j)=
\left\{
\begin{array}{cc}
M(v_j) & j\neq i \\
M'(v_j)& j = i
\end{array}
\right.
\]
It can be easily checked that:
\begin{itemize}
\item $v_i$ is fairly classified by $\tilde{M_i}$
\item By construction $f_{\theta,\Delta}(V,\tilde{M_i}) \geq f_{\theta,\Delta}(V,M)$
\item $\tilde{\varepsilon}_i\geq\varepsilon$
\end{itemize}
where $\varepsilon$ denotes the mean square error of a given outcome $M$. \\
On the other hand, by simply adding a positive quantity we have:
$$\varepsilon\leq\tilde{\varepsilon}_i\leq\varepsilon+\frac{(M'(v_i)-\hat{y}_i)^2}{p}.$$
By iteration adding fair evaluation to the model we obtain an outcome $\tilde{M}$ such that:
$$\varepsilon\leq\tilde{\varepsilon}\leq\varepsilon+\frac{max_{\Delta}(M'(v_j)-\hat{y}_j)^2}{p} \cdot card(\Delta)$$

It is advantageous to use a model that considers all variables and excludes them based on objective criteria, such as error optimization obtained on a statistical basis (e.g., Mean Squared Error - MSE). If optimization excludes variables that might be considered sensitive, they would either have a low predictive impact or induce bias, and therefore would not statistically significantly affect the model's decisions or its fairness, as highlighted by Verma and Rubin (2018) in their analysis of fairness definitions \cite{verma2018}. Subsequent fairness tests operate on an optimized set of variables, with a reduced risk of implicit correlation, aligning with the insights of González-Sendino et al. (2023), who review approaches to bias and fairness in AI \cite{gonzalez2023}. This approach reduces biases, as noted by Žliobaitė (2017) who underscores the importance of measuring discrimination to ensure fairness in algorithmic decisions \cite{zliobaite2017}.

However, despite these mathematical strategies, our exploration shows that excluding critical variables in AI models can lead to a significant loss of fairness and explainability. This highlights the complexity of achieving true fairness in AI systems and underscores the need to integrate ethical considerations with mathematical approaches.
Therefore, it is imperative to adopt a more comprehensive strategy that balances technical methodologies with ethical imperatives. By doing so, we can develop AI systems that are not only mathematically sound but also ethically justifiable and transparent, ultimately upholding human values and fostering trust in AI technologies.\\

\section{Conclusion} \label{sec:Conclusion}

In conclusion, our exploration demonstrates that excluding critical variables in AI models leads to a potential loss of fairness and explainability. Although mathematical measures offer essential tools for assessing algorithmic fairness~\cite{verma2018}, they do not fully encapsulate the philosophical and sociological essence of what fairness means from a human and legal perspective~\cite{weinberg2022}. AI models, constrained by the datasets they are trained in, represent only a limited subset of reality, not capturing the complexity of human society~\cite{gonzalez2023,zliobaite2017}.

This limitation highlights the notion that models are only valid within the scope of their data analysis, and therefore making fair decisions requires considering more than just variables and models; it requires accounting for social implications~\cite{benthall2019,foulds2020}. As Li et al.~\cite{li2022} emphasize, achieving true fairness in AI goes beyond technical solutions; it involves integrating philosophical inquiry and ethical considerations into the development process.

Our findings align with the objectives of the European Union's AI Act~\cite{EUAIAct2024}, which seeks to promote trustworthy and fair AI systems. However, while the AI Act addresses fairness from a regulatory and technical point of view, it may not fully encompass the deeper philosophical aspects of fairness~\cite{varona2022}. Therefore, we advocate for a holistic approach to AI fairness, one that merges mathematical rigor with philosophical ethics and social awareness.

By adopting this interdisciplinary perspective, we can better ensure that AI systems not only perform effectively within their intended data domains, but also contribute positively to society at large~\cite{ruf2021}. This approach resonates with calls for collaboration between technologists, ethicists, and social scientists to address fairness in AI in a comprehensive way~\cite{chen2023}. Ultimately, striving towards an AI that is both fair and explainable requires us to engage with the broader ethical and societal contexts in which these technologies operate.

\section{Future Research} \label{sec:FuturerResearch}

Future work could investigate the integration of philosophical theories of fairness and comparative legal frameworks into the design and assessment of AI systems. Moreover, mathematical models should not only based their reasoning on the mathematical model fairness but also emphasize empirical studies from difference science that help to develop tools and methodologies that not only quantify algorithmic fairness but also provide qualitative analysis of social impacts, unintended consequences, and trade-offs, moving beyond technical audits. Finally, observe how fairness will evolve over time by conducting long-term studies to evaluate how AI influences social outcomes, trust, and perceptions, with the aim of identifying best practices and areas for improvement.

\section*{Acknowledgments}
This work has been supported by Project ECS 0000024 Rome Technopole - CUP B83C22002820006, “PNRR Missione 4 Componente 2 Investimento 1.5,” funded by the European Commission - NextGenerationEU.

\bibliography{PaperBib}

\end{document}